\documentclass[11pt]{article}
\newcommand{\bc}{\begin{center}}
\newcommand{\ec}{\end{center}}
\newcommand{\be}{\begin{equation}}
\newcommand{\ee}{\end{equation}}
\newcommand{\bea}{\begin{eqnarray}}
\newcommand{\eea}{\end{eqnarray}}
\newcommand{\ba}{\begin{array}}
\newcommand{\ea}{\end{array}}
\newcommand{\edc}{\end{document}}

\def\g{\gamma}

\def\O{\Omega}

\def\t{\theta}
\def\b{\beta}

\def\s{\sigma}
\def\m{\mu}

\def\Z{\mathbf Z}

\def\1{\mathbf 1}

\def\L{\Lambda}

\begin{document}
\thispagestyle{empty}

{\Large {\bf On a Phase Separation Point for One - Dimensional
Models}} \vspace{0.3cm}
\begin{center}
{\bf N.N. Ganikhodjaev$^{1,2}$, U.A. Rozikov$^{1,3}$}\\
 $^1$Institute of Math. and Infor. Technol., 29, F.Hodjaev str., 100125, Tashkent,
 Uzbekistan.\\
$^2$International Islamic University Malaysia, P.O. Box 141, 25710, Kuantan, Malaysia.\\
$^3$ School of Math. Sci. GC University, Lahore, Pakistan.\\
 E-mail: nasirgani@yandex.ru, rozikovu@yandex.ru\\[2mm]
\end{center}
\vspace{0.4cm}

{\bf Abstract:} In the paper a one-dimensional model with nearest
- neighbor interactions $I_n, n\in \Z$ and  spin values $\pm 1$ is
considered. It is known that under some conditions on parameters
$I_n$  the phase transition occurs for the model. We define a
notion of "phase separation" point between two phases. We prove
that the expectation value of the point is zero and its the mean
square fluctuation is bounded by a constant $C(\beta)$
which tends to $\frac{1}{4}$ if $\beta\to\infty$.
Here $\beta=\frac{1}{T}$, $ T>0$-temperature. \\[1mm]

\section{Introduction}

It is known that the curve of separation between two pure phases
for two dimensional Ising model at low temperature is non rigid:
Gallavotti [8] showed that the mean square fluctuation of the
height of the interface, or phase separation line, diverges as
$\sqrt{L}$ (where $L$ is side of square) in the thermodynamic
limit. A different picture occurs in the three-dimensional case:
there exists a value $\b_{\rm r}>\b_{\rm cr}$ (where $\b_{\rm cr}$
is the critical value of inverse temperature $\b$ of phase
transition for the model) such that for $\b>\b_{\rm r}$ the phase
separation membrane is at a finite distance from the plane (x=0).
For values of $\b$ between $\b_{\rm cr}$ and $\b_{\rm r}$ the
membrane deviates from the plane (x=0) at the distance $\sim \log
L.$ (see [1], [4], [13]). These results were obtained for the
model with short range, translation invariant interactions. In
[18] Van Hove showed (see also [16, section 5.6.]) that a
one-dimensional system could not exhibit a phase transition if the
(translation-invariant) forces were of finite range. However, by
breaking translation invariance we can obtain a phase transition
in one-dimensional models with only nearest neighbor interactions
[17], [10, p.95], [15]. In [3], [5-7], [10]-[12] other examples of
phase transitions were considered for one-dimensional models with
long range interactions.

In the paper we consider the Hamiltonian
$$H(\s)= \sum_{l=(x-1, x): x\in Z}
I_x\1_{\s(x-1)\ne\s(x)}, \eqno(1)$$ where $Z=\{...,-1,-2, 0,
1,2,...\}$, $\s=\{\s(x)\in \{-1,1\}: x\in Z\} \in \O=\{-1,1\}^Z,$
and $ I_x\in R$ for any $x\in Z.$

Note that [10, p.95] for the model (1) on $N=\{1,2,...\}$ it was
shown that there occurs a phase transition iff $\sum_{n\geq
1}e^{-2I_n}<\infty.$ In [15] using a contour argument it has been
proven that for that model (1) the phase transition occurs if
$I_n+I_{n+k}>k$ for any $n\in Z, k\in N.$

In two (resp. three) dimensional case the phase separation curve
(resp. membrane) is defined as an "open" contour [4],[8]. But that
construction does not work for one dimensional case when
interactions are only  nearest neighbors. In this case the
separation "line" is a point.  To the best of our knowledge, there
is no any paper devoted to PSP of one - dimensional models. One of
the main reasons of this absence, we think,  can be the fact that
a one - dimensional model has phase transition if ether it has
long range (Dyson model) or non translational-invariant
(Sullivan's model) interactions. Therefore for such models
definition and investigation of PSP is rather difficult problem.

In the present paper we are going to give a more natural
definition of the phase separation point (PSP) between two phases
in one dimensional setting. For the model (1) we show that the
expectation value of the point is zero and its the mean square
fluctuation (for $\beta>\beta_c$) is bounded by a constant
$C(\beta)$ which tends to $\frac{1}{4}$ if $\beta\to\infty$. Thus
in one - dimensional case the behavior  of the PSP is very
different from behavior of the PS line (membrane) of two (three)
dimensional Ising model mentioned above.

\section{ "+" boundary condition}
Let us consider a sequence $\L_n=[-n,n], n=0,1,...$ and denote
$\L_n^c=Z\setminus \L_n$. Consider a boundary condition
$\s^{(+)}_n=\s_{\L_n^c}=\{\s(x)=+ 1: x\in \L_n^c\}.$ The energy
$H_n^+(\s)$ of the configuration $\s$ in the presence of the
boundary condition $\s^{(+)}_n$ is expressed by the formula
$$H^+_n(\s)= \sum_{l=(x-1, x): x\in \L_n}
I_x\1_{\s(x-1)\ne\s(x)}+
I_{-n}\1_{\s(-n)\ne 1}+I_{n+1}\1_{\s(n)\ne 1}. \eqno(2)$$

The Gibbs measure on $\O_n=\{-1, 1\}^{\L_n}$ with respect to the
boundary condition $\s_n^{(+)}$ is defined by the usual way

$$\m^+_{n,\b}(\s)=Z^{-1}(n,\b,+)
\exp(-\b H_n^+(\s)), \eqno(3)$$
where $\b=T^{-1}$, $T>0-$ temperature and
 $ Z(n, \b, +)$ is the normalizing factor (statistical sum).

Denote by $\s^+_n$ the configuration on $Z$ such that
$\s^+_n(x)=+1$ for any $x\in \L_n^c.$

Put
$$A(\s^+_n)=\{x\in Z: \s^+_n(x)=-1\}.$$
Note that there is a one-to-one correspondence between the set of
all configurations $\s^+_n$ and the set of all subsets of $\L_n.$

Let $A'(\s^+_n)$ be the set of all maximal connected subsets of
$A(\s^+_n).$

\vspace{0.3cm}
 {\bf Lemma 1}.[15] {\it Let $B\subset Z$ be a fixed connected set and
$p_\b^+(B)=\mu^+_{n,\b}\{\s^+_n: B\in A'(\s^+_n)\}.$ Then
$$p^+_\b(B)\leq \exp\bigg\{-\b\bigg[I_{n_B}+I_{N_B+1}\bigg]\bigg\}, $$
where $n_B$ (resp. $N_B$) is the left (resp. right)  endpoint of $B.$}
\vspace{0.3cm}

Assume that the coupling interactions of the Hamiltonian (1)
satisfy the following condition
$$ I_n+I_{n+r}\geq r  \ \ \textrm{for any} \ r\in \{1,2,...\} \
\textrm{and} \ n\in Z\eqno(4)$$

\vspace{0.3cm}
 {\bf Theorem 2.} [15] {\it Assume the condition (4) is satisfied.
 For all sufficiently large $\b$ there are at least
two Gibbs measures for the  model (1).}

\vspace{0.3cm}

Denote
$${\cal H}=\{H: H \mbox{ (see (1)) satisfies the condition (4)}\}$$
The following example shows that the set ${\cal H}$ is not empty.

\vspace{0.3cm} {\bf Example.} Consider Hamiltonian  (1) with
$I_m\geq |m|,\ \  m\in Z.$ Then
$$I_m+I_{m+k}\geq |m|+|m+k|\geq k$$
for all $m\in Z$ and $k\geq 1.$ Thus the condition (4) is
satisfied.

\section{"$\pm$" -boundary condition}

\subsection{ Statistical sum}

Consider two type of statistical sums:

$$Z^+_n=\sum_{\s_n\in \O_n}\exp\{-\b H^+_n(\s_n)\}, \eqno(5)$$
$$Z^{\pm}_n=\sum_{\s_n\in \O_n}\exp\{-\b H^{\pm}_n(\s_n)\}, \eqno(6)$$
where $H^+_n$ is defined by (2) and
$$H^{\pm}_n(\s_n)=H^+_n(\s_n)+I_{-n}\s(-n). \eqno (7)$$
In this paper for the simplicity assume
$$I_n=I_{-n+1},\ \ \mbox{for any} \ \ n\in Z. \eqno (8)$$
Under the condition (8) we get
$$Z^-_n=Z^+_n \ \ \mbox{and} \ \ Z^{\pm}_n=Z^{\mp}_n. \eqno(9)$$
Using (8) and (9) from (5),(6) we obtain
$$\begin{array}{llll}
Z^+_n=(1+e^{-2\b I_{n+1}})Z^+_{n-1}+2e^{-\b
I_{n+1}}Z^{\pm}_{n-1},\\
 Z^{\pm}_n=(1+e^{-2\b
I_{n+1}})Z^{\pm}_{n-1}+2e^{-\b I_{n+1}}Z^+_{n-1}
\end{array}\eqno(10)$$

Putting $X_n=Z^+_n-Z^{\pm}_n$ and $Y_n=Z^+_n+Z^{\pm}_n$ from (10)
one gets
$$\begin{array}{llll}
X_n=(1-e^{-\b I_{n+1}})^2X_{n-1},\\
Y_n=(1+e^{-\b I_{n+1}})^2Y_{n-1}.\\
\end{array}\eqno(11)$$

The equalities $X_0=Z^+_0-Z^{\pm}_0=(1-e^{-\b I_1})^2, \ \
Y_0=(1+e^{-\b I_1})^2$  with (11) imply
$$
X_n=\prod_{i=0}^n(1-e^{-\b I_{i+1}})^2,\\
Y_n=\prod_{i=0}^n(1+e^{-\b I_{i+1}})^2.\\$$ Hence
$$\begin{array}{llll}
Z_n^+={1\over 2}\bigg(\prod_{i=0}^n(1+e^{-\b I_{i+1}})^2+\prod_{i=0}^n(1-e^{-\b I_{i+1}})^2\bigg),\\
Z_n^{\pm}={1\over 2}\bigg(\prod_{i=0}^n(1+e^{-\b I_{i+1}})^2-\prod_{i=0}^n(1-e^{-\b I_{i+1}})^2\bigg).\\
\end{array}\eqno(12)$$

For example, in a case of the usual Ising model i.e. $I_n=I,$
$\forall n$ from (12) denoting $\tau=\exp(-\b I)$ we get
$$\begin{array}{llll}
Z_n^+={1\over 2}\bigg((1+\tau)^{2(n+1)}+(1-\tau)^{2(n+1)}\bigg),\\
Z_n^{\pm}={1\over 2}\bigg((1+\tau)^{2(n+1)}-(1-\tau)^{2(n+1)}\bigg).\\
\end{array}$$ Using these equalities (for usual Ising model) it is easy to see that
$$ {Z_n^+\over Z_n^{\pm}}\to 1, \ \ {\rm if} \ \ n\to\infty .$$

\subsection{Phase-separation point}

Fix $n\in \{0,1,2,...\}.$ Denote by $\O_n$ the set of all
configurations on $\L_n=\{-n,...,n\}$ i.e. $\O_n=\{-1,1\}^{\L_n}.$
For every $\s_n\in \O_n$ define $\s^{\pm}_n\in \{-1,1\}^Z$ as
follows
$$ \s^{\pm}_n(x)=\left\{
\begin{array}{lll}
-1 & \textrm{if $x<-n$}\\
\s_n(x) & \textrm{if $x\in \L_n$}\\
1 & \textrm{if $x>n$}\\
\end{array}\right. \ \ \ \ x\in Z. \eqno(13)$$
Let $\O^{\pm}_n$ be the set of all configurations defined by (13).
Denote
$$\O^{(+)}_n=\{\s_n\in \O^{\pm}_n:
|\{x\in \L_n: \s_n(x)=1\}|\geq n+1\};$$
$$\O^{(-)}_n=\{-\s_n: \s_n\in \O^{(+)}_n\}.$$
Clearly $\O^{(+)}_n\cap\O^{(-)}_n=\emptyset$ and
$\O^{\pm}_n=\O^{(+)}_n\cup \O^{(-)}_n.$

Let $S:\O^{\pm}_n\to \O^{\pm}_n$ be operator such that
$$ S(\s_n)(x)=-\s_n(-x), \ \ x\in Z, \eqno(14)$$
It is easy to see that
$$S(\O^{(\pm)}_n)=\O^{(\mp)}_n, \eqno(15)$$
i.e. the operator $S$ is one-to-one map from $\O^{(+)}_n$ (resp.
$\O^{(-)}_n$) to $\O^{(-)}_n$ (resp. $\O^{(+)}_n$).

\vskip 0.2 truecm

{\bf Lemma 3.} {\it The Hamiltonian (1) (under condition (8)) is
invariant with respect to operator $S$ i.e. $H(S(\s))=H(\s)$ for
any} $\s\in \O^{\pm}_n, n=0,1,...$

\vskip 0.2 truecm

 {\bf Proof.} Note that operator $S$ is the
combination of the following two symmetry maps $U:\O^{\pm}_n\to
\O^{\pm}_n$  such that $ U(\s_n)(x)=-\s_n(x),$ and
$V:\O^{\pm}_n\to \O^{\pm}_n$  such that $ V(\s_n)(x)=\s_n(-x).$
Clearly, $H$ is invariant with respect to $U$ and $V$ this
completes the proof.

Denote $T_n=\{-n-{1\over 2}, -n+{1\over 2},...,n-{1\over
2},n+{1\over 2}\}.$ Fix $\s_n\in \O^{\pm}_n$ and we say that $t\in
T_n$ is an interface point for the configuration $\s_n$ if
$\s_n(t-{1\over 2})\ne \s_n(t+{1\over 2}).$ For any interface
point $t\in T_n$ denote
$$l^-_t\equiv l^-_t(\s_n)=|\{x\in \L_n: \s_n(x)=-1, x<t\}|,$$
$$ r^+_t\equiv r^+_t(\s_n)=|\{x\in \L_n: \s_n(x)=1, x>t\}|,$$
$$ l^+_t=n+t+{1\over 2}- l^-_t, \ \ r^-_t=n-t+{1\over 2}-
r^+_t.$$
$$\Delta_t=(l^-_t,r^+_t), \ \ \|\Delta_t\|=l^-_t+r^+_t.$$

{\bf Definition 4.} We define  PSP $\g_n(\s_n)\in T_n$ as the
following interface point
$$\g_n(\s_n)=\left\{ \begin{array}{llll}
\max\{t_0\in T_n: \|\Delta_{t_0}\|=\max_t\|\Delta_t\|\}, \ \
\mbox{if} \ \ \s_n\in \O^-_n,\\[3mm]
\min\{t_0\in T_n: \|\Delta_{t_0}\|=\max_t\|\Delta_t\|\}, \ \
\mbox{if} \ \ \s_n\in \O^+_n.\\
\end{array}\right. \eqno(16)$$

\vskip 0.2 truecm

{\bf Lemma 5.} {\it For any $\s_n\in \O^{\pm}_n$ we have
$$ \g_n(\s_n)=-\g_n(S(\s_n)). \eqno(17)$$}

\vskip 0.1 truecm {\bf Proof.} Straightforward.

For $\t\in T_n$ denote
$$P_n(\t)=\mu^{\pm}_n\{\s_n: \g_n(\s_n)=\t\},$$
where $\mu^{\pm}_n$ is the Gibbs measure with respect to
$\pm$-boundary condition.

\vskip 0.2 truecm

{\bf Lemma 6.} {\it For any $\t$ and $n\in N$ we have
$$P_n(\t)=P_n(-\t).$$}
\vskip 0.2 truecm {\bf Proof.} The proof follows from lemma 3, and
equality (17).

\vskip 0.2 truecm

As a corollary of lemmas 3 and 6 we have

\vskip 0.2 truecm {\bf Lemma 7.} {\it For any $n\in N$
$${\mathbf E}_{\mu^{\pm}_n}(\g_n)=0,$$
where ${\mathbf E}_{\mu^{\pm}_n}$ is the expectation value of the
random variable $\g_n$ with respect to the Gibbs measure
$\mu^{\pm}_n.$}

For a given configuration $\s_n$ denote by  $\t_1<\t_2<...<\t_k$
the interface points generated by $\s_n$.
 \vskip 0.2 truecm
 {\bf Theorem 8.} {\it 1. If an interface point
$t=\t_1$, ({\rm resp.}\ \  $t=\t_k$) is  PSP then
$$l^-_t\geq l^+_t=0,\ \ r^+_t>r^-_t,\ \
({\rm resp.}\ \  l^-_t>l^+_t,\ \ r^+_t\geq r^-_t=0).\eqno (18)$$

2. If an interface point $t\in T_n, t\ne \t_1, \t_k$ is  PSP then}
$$l^-_t>l^+_t,\ \ r^+_t>r^-_t.\eqno(19)$$

{\bf Proof.} Consider case $\s_n\in \O^{(+)}_n$ and $t\ne \t_1,
\t_k$  (all other cases can be proved similarly). Assume
$l_t^-\leq l^+_t$ then
$$\|\Delta_t\|=l^-_t+r^+_t<l^+_t+r^+_t=r^+_{\t_1}\leq
\|\Delta_{\t_1}\|.$$  Thus by definition we get $\g_n(\s_n)=\t_1,$
which contradicts to $t\neq \t_1$. This completes the proof.

{\bf Remark 1.} In general for a given configuration $\s_n$ a
point $t$ satisfying the condition (18),(19) is not unique. For
example, take $\s_2=\{\s_2(-2)=-1, \s_2(-1)=-1, \s_2(0)=1,
\s_2(1)=-1, \s_2(2)=1\},$ the interface points $t=-0.5$ and
$t=1.5$ satisfy the condition (19).  Thus the conditions (18),(19)
are necessary for $t$ to be PSP but are not sufficient.

Summing over all configurations with a given $\t$ we obtain the
probability $P_n(\t)$ of $\t$ which can be written by
$$P_n(\t)={e^{-\b I_{\t+1/2}} Y_{-n, \t-1/2}Y_{\t+1/2, n}\over
Z^{\pm}_n}, \eqno(20)$$ where $Z^{\pm}_n$ is defined by (12) and $
Y_{-n, \t-1/2}$ (resp. $Y_{\t+1/2, n}$)  is the "crystal"
partition function which contains {\it only} sum of terms
$\exp(-\b H^+(\varphi))$ with $\varphi=\s'\in \{-1,1\}^{[-n,
\t-3/2]}$ (resp. $\varphi=\s"\in \{-1,1\}^{[\t+3/2, n]}$) such
that the PSP of the total configuration
$\s=\s'\cup\{\s(\t-1/2)=-1,\ \  \s(\t+1/2)=1\}\cup \s"$ on
$[-n,n]$ is $\t$.

{\bf Remark 2.} In two-dimensional Ising model case an analog of
the formula (20) is given in [2, formula (3.2)]. Comparing our
formula (20) with the formula (3.2) we notice that the numerator
of the formula (3.2) contains a product of "full" (all possible
terms) partition functions with pure $"+"$ boundary conditions (or
$"-"$ boundary conditions which is equivalent by symmetry) in the
different connected components of $Z^2$ which are separated by the
phase separation curve. But in our setting the numerator of the
formula (20) contains product of {\it crystal} partition functions
which we have defined above. This is a remarkable difference
between the notions of phase separation of one and two dimensional
Ising models. In the sequel of this section we are going to
estimate the crystal partition functions by "rarefied"  partition
functions.

By Lemma 6 it is enough to consider the case $\t\geq \frac{1}{2}$.
For $A\subset \Z$ we denote  $\O_A=\{-1,1\}^A$ -the set of all
configurations defined on $A$. Denote
$$H^-_{n,\t}(\s)=\sum^{\t-\frac{3}{2}}_{x=-n}I_x\1_{\s(x)\ne\s(x+1)}+
I_{-n}\1_{\s(-n)\ne
-1}+I_{\t-\frac{1}{2}}\1_{\s(\t-\frac{3}{2})\ne -1}, \ \ \s\in
\O_{\{-n,...,\t-\frac{3}{2}\}};$$
$$H^+_{n,\t}(\s)=\sum_{x=\t+\frac{3}{2}}^{n}I_x\1_{\s(x)\ne\s(x+1)}+
I_{n+1}\1_{\s(n)\ne 1}+I_{\t+\frac{3}{2}}\1_{\s(\t+\frac{3}{2})\ne
1}, \ \ \s\in \O_{\{\t+\frac{3}{2},...,n\}};$$
$$H^{\pm}_{n,\t}(\s)=H^-_{n,\t}(\s)-I_{-n}\s(-n);$$
$$H^{\mp}_{n,\t}(\s)=H^+_{n,\t}(\s)+I_{n+1}\s(n).$$
Now we are ready to define the "rarefied" partition functions i.e.
$$\overrightarrow{Z}_{n,\t}=\sum_{\s\in
\O_{\{\t+\frac{3}{2},...,n\}}}\exp(-\b H^+_{n,\t}(\s));\ \
\overrightarrow{Z}_{n,\t}^-=\sum_{\s\in
\O_{\{\t+\frac{3}{2},...,n\}}}\exp(-\b H^{\mp}_{n,\t}(\s));$$
$$\overleftarrow{Z}_{n,\t}=\sum_{\s\in
\O_{\{-n,...,\t-\frac{3}{2}\}}}\exp(-\b H^-_{n,\t}(\s));\  \
\overleftarrow{Z}_{n,\t}^+=\sum_{\s\in
\O_{\{-n,...,\t-\frac{3}{2}\}}}\exp(-\b H^{\pm}_{n,\t}(\s)).$$

Note that (see (20))
$$Y_{-n, \t-1/2}\leq \overleftarrow{Z}_{n,\t}; \ \ Y_{\t+1/2, n}\leq
\overrightarrow{Z}_{n,\t}. \eqno (21)$$

It is easy to check that
$$\begin{array}{llllll}
\overrightarrow{Z}_{n,\t}=\overrightarrow{Z}_{n-1,\t}+e^{-\b
I_{n+1}}\overrightarrow{Z}^-_{n-1,\t},\ \ n\geq
\t+\frac{3}{2}\\[3mm]
\overrightarrow{Z}^-_{n,\t}=\overrightarrow{Z}_{n-1,\t}^-+e^{-\b
I_{n+1}}\overrightarrow{Z}_{n-1,\t},\\[3mm]
\overrightarrow{Z}_{\t+\frac{1}{2}, \t}=1; \ \
\overrightarrow{Z}^-_{\t+\frac{1}{2},\t}=e^{-\b
I_{\t+\frac{3}{2}}}.
 \end{array}\eqno(22)$$
 Denote
 $u_{n,\t}=\overrightarrow{Z}_{n,\t}-\overrightarrow{Z}_{n,\t}^-$,
 $v_{n,\t}=\overrightarrow{Z}_{n,\t}+\overrightarrow{Z}_{n,\t}^-$.
 Then from (22) we get
$$\begin{array}{llllll}
u_{n,\t}=\left(1-e^{-\b I_{n+1}}\right)u_{n-1,\t},\ \ n\geq
\t+\frac{3}{2},\\[3mm]
u_{\t+\frac{1}{2}, \t}=1-e^{-\b I_{\t+\frac{3}{2}}},
 \end{array}$$ i.e.
$$u_{n,\t}=\prod_{i=\t+\frac{1}{2}}^n\left(1-e^{-\b I_{i+1}}\right).$$
Similarly $$v_{n,\t}=\prod_{i=\t+\frac{1}{2}}^n\left(1+e^{-\b
I_{i+1}}\right).$$ Hence

$$\begin{array}{llllll}
\overrightarrow{Z}_{n,\t}=\frac{1}{2}\left(\prod_{i=\t+\frac{1}{2}}^n(1+e^{-\b
I_{i+1}})+\prod_{i=\t+\frac{1}{2}}^n(1-e^{-\b
I_{i+1}})\right);\\[3mm]
\overrightarrow{Z}_{n,\t}^-=\frac{1}{2}\left(\prod_{i=\t+\frac{1}{2}}^n(1+e^{-\b
I_{i+1}})-\prod_{i=\t+\frac{1}{2}}^n(1-e^{-\b I_{i+1}})\right).
 \end{array}\eqno(23)$$

 Analogically, using condition (8) we get

$$
\overleftarrow{Z}_{n,\t}=\frac{1}{2}\left(\prod_{i=\t+\frac{3}{2}}^n(1+e^{-\b
I_{i+1}})\prod^{\t-\frac{3}{2}}_{i=1}(1+e^{-\b I_{i+1}})^2
+\prod_{i=\t+\frac{3}{2}}^n(1-e^{-\b
I_{i+1}})\prod^{\t-\frac{3}{2}}_{i=1}(1-e^{-\b I_{i+1}})^2
\right); \eqno (24)$$
$$
\overleftarrow{Z}_{n,\t}^+=\frac{1}{2}\left(\prod_{i=\t+\frac{3}{2}}^n(1+e^{-\b
I_{i+1}})\prod^{\t-\frac{3}{2}}_{i=1}(1+e^{-\b I_{i+1}})^2
-\prod_{i=\t+\frac{3}{2}}^n(1-e^{-\b
I_{i+1}})\prod^{\t-\frac{3}{2}}_{i=1}(1-e^{-\b I_{i+1}})^2
\right).
$$
Using formulas (12), (23), (24) and inequalities (21) from (20)
one gets a upper bound of $P_n(\t)$.

\section{Variation of the PSP}

In this section, for simplicity, we consider the following case
$$I_n=\left\{\begin{array}{ll}
n, \ \ \mbox{if} \ \ n>0\\
-n+1, \ \ \mbox{if} \ \ n\leq 0
\end{array}\right. \eqno(25)$$
By lemmas 6 and 7 the variation of $\g_n$ can be written as
$${\rm
Var}(\g_n)=2\sum_{\t=\frac{1}{2}}^{n+\frac{1}{2}}\t^2P_n(\t).\eqno(26)$$

{\bf Theorem 9.} {\it If interactions $I_n$ satisfy (25) and
$\beta$ large enough then
$$\frac{1}{4}\leq {\rm
Var}(\g_n) \leq\sim \frac{\tau
A(\tau)\cosh\left(\frac{\tau^2}{1-\tau}\right)}{2\sinh(2\tau)}\left(1+\frac{3\tau(\tau+3)}{(1-\tau)^2}\right),$$
where $\tau=e^{-\beta}$,} $A(\tau)=
\cosh\left(\frac{\tau^2}{1-\tau}\right)\cosh\left(\tau(1+\tau)\right)-\sinh\left(\tau^2\right)
\sinh\left(\tau(1+\tau)\right)$.

{\bf Proof.} The lower bound easily follows from (26). We shall
prove upper bound. It follows from (26), (20) and (21) that
$${\rm
Var}(\g_n)=2\sum_{\t=\frac{1}{2}}^{n+\frac{1}{2}}\t^2 {e^{-\b
I_{\t+1/2}} Y_{-n, \t-1/2}Y_{\t+1/2, n}\over Z^{\pm}_n}\leq$$
$$2\sum_{\t=\frac{1}{2}}^{n+\frac{1}{2}}\t^2
{e^{-\b I_{\t+1/2}} \overleftarrow{Z}_{n,
\t}\overrightarrow{Z}_{n,\t}\over Z^{\pm}_n}
$$
 By (25) from (12) we get
$$Z^{\pm}_n=\frac{1}{2}\left(\exp\left(2\sum_{i=0}^n\ln(1+\tau^{i+1})\right)-
\exp\left(2\sum_{i=0}^n\ln(1-\tau^{i+1})\right)\right)\sim$$
$$
\frac{1}{2}\left(\exp\left(2\sum_{i=0}^n\tau^{i+1}\right)-\exp\left(-2\sum_{i=0}^n\tau^{i+1}\right)\right)=
\sinh\left(\frac{2\tau(1-\tau^{n+1})}{1-\tau}\right)\geq
\sinh(2\tau).\eqno(27)$$ Here we used $\ln(1+\tau^i)\sim \tau^i$
for small $\tau$ (i.e. large $\beta$).

Similarly from (23) and (24) for $\t\geq \frac{1}{2}$ we get
$$\overrightarrow{Z}_{n,\t}\sim\cosh\left(\frac{\tau^{\t+\frac{3}{2}}
\left(1-\tau^{n-\t+\frac{1}{2}}\right)}{1-\tau}\right)\leq
\cosh\left(\frac{\tau^{\t+\frac{3}{2}}}{1-\tau}\right)\leq
\cosh\left(\frac{\tau^2}{1-\tau}\right); \eqno(28)$$
$$\overleftarrow{Z}_{n,\t}\sim\cosh\left(\frac{\tau^2
\left(1-\tau^n\right)}{1-\tau}-\tau^{\t+\frac{1}{2}}(1+\tau)\right)\leq
A(\tau). \eqno(29)$$ Hence
$${\rm Var}(\g_n)\leq 2\tau\left(\frac{1}{4}+\sum^\infty_{m=1}(m+\frac{1}{2})^2\tau^m
\right)\frac{\cosh\left(\frac{\tau^2}{1-\tau}\right)A(\tau)}{\sinh(2\tau)}.
\eqno(30)$$ One can check that
$$\sum^\infty_{m=1}(m+\frac{1}{2})^2\tau^m=\frac{3\tau(\tau+3)}{4(1-\tau)^2}.\eqno (31)$$
Thus from (30) and (31) one gets the assertion of the Theorem.

{\bf Remark 3.} The estimation $\frac{1}{4}\leq {\rm Var}(\g_n)$
is true for any interactions $I_n$, i.e. the condition (25) is not
necessary.

{\bf Corollary.} {\it For any $n$ the following holds}
$$\lim_{\b\to\infty}{\rm Var}(\g_n)=\frac{1}{4}.$$

\section{Conclusions}

 In usual one-dimensional case
there can be no phase transition. But it is known that phase
transition occurs in the following cases:

a) The set of spin values is $\{-1,1\}$ and interactions are long
range (Dyson's model).

b) The set of spin values is $\{-1,1\}$ and interactions are
nearest neighbors, but they are spatially inhomogeneous
(Sullivan's model).

c) The set of spin values is a countably infinite set and
interactions are nearest neighbors (Spitzer's model).

For a detailed description of history of phase separation
properties of lattice models see [9].

We have considered here a one-dimensional model of type b). As
mentioned above for such kind of model a phase transition occurs
(see Theorem 2). In such a case of the  existence of the phase
transition, it would be interesting to know certain properties of
a phase separation point (PSP) (curve (membrane) in two (three)
dimensional case).

In two (resp. three) dimensional case a phase separation curve
(resp. membrane) is defined as an "open" contour [4],[8]. But this
construction does not work for one dimensional case with
interactions of only  nearest neighbors. A notion of PSP, to our
knowledge, have not yet been introduced for one-dimensional
models. In one-dimensional case the separation "line" is a point.
We have introduced here a natural definition of the PSP between
two phases in one-dimensional case.  We studied asymptotical
properties of the PSP. Our definition of PSP is rather natural and
properties of the PSP more special than ones in two and three
dimensional cases. Namely, from Theorem 9 it follows that with
probability 1 the PSP should be $-\frac{1}{2}$ or $\frac{1}{2}$.

 \vspace{0.2cm} {\bf
Acknowledgments.} A part of this work was done within the scheme
of Junior Associate at the ICTP, Trieste, Italy and RUA thanks
ICTP for providing financial support and all facilities.  The work
partically supported by the SAGA Fund P77c
 of the Ministry of Science, Technology and Innovation (MOSTI) through the
 Academy of Sciences Malaysia. RUA thanks MOSTI and IIUM,
 for support and hospitality (in July-August 2007).

We thank A.C.D. van Enter who gave us useful information about the
Ising chain on $N$ with inhomogeneous interactions.
 We also gratitude to
F.Mukhamedov for many helpful discussions and C.H.Pah for some
numerical analysis which were useful to understand our results.

\vskip 0.2 truecm {\bf References}

1. Abraham D.B. Surface structures and phase-transitions--exact
results. Phase transitions and Critical Phenomena, V.10, C.Domb
and J.Lebowits (eds.), Acad. Press, New York and London, 1986,
p.1-69.

2. Bricmont J., Lebowitz J. and Pfister C. On the local structure
of the phase separation line in the two-dimensional Ising system.
{\it Jour. Stat. Phys.} {\bf 26}: 313-332 (1981).

3. Cassandro M., Ferrari P., Merola I., Presutti E. Geometry of
contours and Peierls estimates in $d=1$ Ising models with long
range interactions. {\it J. Math.Phys.} {\bf 46}: 053305-053327
(2005).

4. Dobrushin R. Gibbs state describing coexistence of phases for a
three-dimensional Ising model. {\it Theor. Prob. Appl.} {\bf 17}:
582-600 (1972).

5. Dyson F. Existence of phase transition in a one - dimensional
Ising ferromagnet. {\it Commun. Math. Phys.} {\bf 12}: 91-107
(1969).

6. Dyson F. An Ising ferromagnetic  with discontinuous long-range
order. {\it Commun. Math. Phys.} {\bf 21}: 269-283 (1971).

7. Flohlich J., Spencer T., The phase transition in the
one-dimensional Ising model with ${1\over r^2}$ interaction
energy. {\it Commun. Math. Phys.} {\bf 84}: 87-101 (1982).

8. Gallavotti G. The phase separation line in the two-dimensional
Ising model. {\it Commun. Math. Phys.} {\bf 27}: 103-136 (1972).

9. Gallavotti G. Statistical mechanics, A short treatise. {\it
Text and Monographs in Physics,} Springer, 1999.

10. Georgii H.-O. Gibbs measures and phase transitions. {\it de
Gruyter stud. in Math.} V.9., Berlin. 1988.

11. Johansson K., Condensation of a one-dimensional lattice gas.
{\it Commun. Math. Phys.} {\bf 141} : 41-61 (1991).

12. Johansson K., On separation of phase in one-dimensional gases.
{\it Commun. Math. Phys.} {\bf 169} : 521-561 (1995).

13. Minlos R. Introduction to mathematical statistical physics.
{\it Univ. Lecture series} V.19. AMS. 2000.

14. Niven I. Mathematics of choice. {\it The Math. Assoc.
America.} V.15. 1965.

15. Rozikov U. An example of one-dimensional phase transition.
{\it Siber. Adv. Math.} {\bf 16}: 121-125 (2006).

16. Ruelle D., {\it Statistical mechanics (rigorous results)},
Benjamin, New York, 1969.

17. Sullivan W. Exponential convergence in dynamic Ising models
with Distinct phases. {\it Phys. Letters} {\bf 53A}: 441-442
(1975).

18. Van Hove L., Sur l'integrale de  configuration pour les
systems de particles a une dimension, {\it Physica} {\bf 16}:
137-143 (1950).

\end{document}